\documentclass{article}
\usepackage[utf8x]{inputenc}
\usepackage{spconf,amsmath,graphicx}

\graphicspath{{figures/}}

\usepackage{cite}
\usepackage{amsmath,amssymb,amsfonts}
\usepackage{algorithmic}
\usepackage{graphicx}
\usepackage{multirow}
\usepackage{tikz}
\usepackage{hyperref}
\usepackage{xcolor}
\usepackage{lipsum}
\usepackage{caption}
\usepackage{siunitx}
\usepackage{commath}
\usepackage{subfig}
\usepackage{pgfplots}
\pgfplotsset{compat=1.17}
\usepackage{mathrsfs}
\usetikzlibrary{arrows}
\usetikzlibrary{positioning}
\usetikzlibrary{3d}
\usepackage{textcomp}
\usepackage{xcolor}
\def\BibTeX{{\rm B\kern-.05em{\sc i\kern-.025em b}\kern-.08em
    T\kern-.1667em\lower.7ex\hbox{E}\kern-.125emX}}

\begin{document}

\definecolor{col_w}{rgb}{0.870588,0.796078,0.776470}
\definecolor{col_wn}{rgb}{0.996078,0.847059,0.364706}
\definecolor{col_out}{rgb}{1,0.5,0}
\definecolor{col_bc}{rgb}{0,0.5,1}
\definecolor{col_conv}{rgb}{0,1,1}
\definecolor{col_out}{rgb}{1,0.5,0}

\title{Multi-scale Deformable Alignment and Content-Adaptive Inference \\ for Flexible-Rate Bi-Directional Video Compression}

\name{M. Akın Yılmaz, O. Ugur Ulas, A. Murat Tekalp\thanks{This work is supported in part by TUBITAK 2247-A Award No.~120C156 and KUIS AI Center funded by Turkish Is Bank. A. M. Tekalp also acknowledges support from Turkish Academy of Sciences (TUBA).} }

\address{Dept. of Electrical \& Electronics Engineering, Koç University, Istanbul, Turkey}

\maketitle

\begin{abstract}
The lack of ability to adapt the motion compensation model to video content is an important limitation of current end-to-end learned video compression models. This paper advances the~state-of-the-art by proposing an adaptive motion-compensation model for end-to-end rate-distortion optimized hierarchical bi-directional video compression. In particular, we propose two novelties: i) a multi-scale deformable alignment scheme at the feature level combined with multi-scale conditional coding, ii) motion-content adaptive inference. In addition, we employ~a~gain unit, which enables a single model to operate at multiple rate-distortion operating points. We also exploit the~gain unit to control bit allocation among intra-coded vs. bi-directionally coded frames by fine tuning corresponding models for truly flexible-rate learned video coding. Experimental results demonstrate state-of-the-art rate-distortion performance exceeding those of all prior art in learned video coding\footnote{The models and instructions to reproduce our results can be found at 
\url{https://github.com/KUIS-AI-Tekalp-Research-Group/video-compression/tree/master/ICIP2023}.}.
\end{abstract}

\begin{keywords}
bi-directional video compression, hierarchical B pictures, end-to-end rate-distortion optimization, content-adaptive inference, flexible-rate coding
\end{keywords}

\section{Introduction}
\label{intro}
\vspace{-8pt}
Video compression technology is in the midst of a transition from the traditional approaches, such as H.265/HEVC~\cite{h265} and H.266/VVC~\cite{vvc}, to deep learning based models.
Instead of relying on hand-crafted models and algorithms, deep learning-based video compression utilizes machine learning principles to learn the statistical properties of consumer video from available video databases to design efficient compression models. 
The biggest advantage of deep learned video compression models is that they are end-to-end optimized for rate-distortion performance. The nonlinear transform to latent variables, temporal alignment in the feature space, quantization, entropy models, and reconstruction models are all optimized using a single rate-distortion loss. Moreover, unlike traditional methods, the optimization can be carried out with respect to any differentiable loss function balancing multiple fidelity and perceptual objectives.

Yet, unlike traditional codecs, learned models lack the~ability to adapt the learned model to given video content, which is the main reason why they cannot outperform classic codecs, such as VVC-VTM~\cite{vtm}. Classic codecs feature several tools such as variable size block partitioning, multiple intra and inter prediction modes, etc. that allow effective adaptation to specific video content.  This paper advances the~state-of-the-art in learned video compression by proposing an adaptive inference model for end-to-end rate-distortion optimized hierarchical bi-directional video compression. In particular, we propose a multi-scale deformable alignment scheme at the feature level combined with multi-scale conditional coding, and motion-content adaptive inference. In addition, we employ~a~gain unit, which enables a single model to operate at multiple rate-distortion operating points. We also exploit the~gain unit to control bit allocation among intra-coded vs. bi-directionally coded frames by fine tuning corresponding models for truly flexible-rate learned video coding.
\vspace{-4pt}

\section{Related work and Contributions}
\label{related} \vspace{-6pt}

\subsection{Low-Delay (Sequential) Video Compression} \label{seqcomp} \vspace{-4pt}

The first deep learned compression models in the literature were designed for sequential coding aiming to replace all components of a traditional sequential video codec with subnetworks, which are jointly optimized. In~\cite{agustsson_scale}, authors propose a scale-space flow model for motion compensation, which allows modeling of motion uncertainty (e.g., occlusions). Later, \cite{elfvc}~proposed a new model that extends the scale-space flow concept by including an in-loop flow predictor and a novel backbone architecture for analysis and synthesis transforms. \cite{ladune2021conditional} proposed a paradigm shift from residual coding to conditional coding, which brings a huge leap in terms of performance. With conditional coding, the model is not bounded to simple subtraction and addition operations for residual compression. The model itself learns a better nonlinear function to fuse and compress motion compensated frame and current frame. \cite{dcvc,ZH_CVPR22} are other recent methods proposed for sequential coding, which combine conditional coding and motion compensation in the feature space. 


\subsection{Random Access (Bi-directional) Video Compression} \vspace{-4pt}
Not many works exist on learned hierarchical video compression, although it is known that classical B-frame coding has superior RD performance compared to sequential video compression. An early work~\cite{hlvc} employs three hierarchical quality layers and a recurrent post-processing network to  enhance the compressed frames. In our concurrent work~\cite{lhbdc}, we perform in-loop bi-directional motion prediction and employ a learned fusion mask to blend forward and backward motion compensated frames to obtain a smoother residual frame for compression. In~\cite{ladune2021conditional}, authors propose a single model to process I, P and B frames using the principal of conditional coding. In our recent work~\cite{flexrate}, we replaced the pre-trained flow estimation model with a latent representation of motion and direct estimation of motion residuals using a compression bottleneck similar to~\cite{agustsson_scale} and introduced frame-level rate control. 
\vspace{-16pt}

\subsection{Continuous Rate Control} \vspace{-4pt}
Early learned image/video compression works train a set of models, one for each discrete R-D operating point. A continuous R-D curve is then interpolated from 6-8 such discrete points. More recently, new approaches have been proposed to achieve continuous rate adaptation to obtain a continuous R-D curve using a single model without extra training. There are various methods proposed for this task aiming not losing much of performance when switching from independent models to a single model. Authors of~\cite{elfvc} learn an embedding for each compression ratio and train the compression network conditioned on the embedding. These embeddings are then interpolated to achieve continuous rate adaptation. In~\cite{flexrate}, the approach  of~\cite{Cui_2021_CVPR} is adopted and a one dimensional re-scaling vector with the size equal to the number of bottleneck channels is learned along with the compression model. During training the authors learn 4 level of vectors for different compression ratios. At inference, they interpolate the vectors and apply them before quantization and entropy coding. 
\vspace{-6pt}

\subsection{Contributions}
\label{ss:contrib} \vspace{-4pt}
We propose extending implicit motion compensation and conditional coding used in sequential video coding to bi-directional B-picture coding within a flexible-rate framework with content-adaptive inference, in contrast to our previous works~\cite{lhbdc, flexrate}, where we employed bi-directional pixel-space warping and residual coding. Our contributions include: \\
1) The proposed novel B-frame compression framework employs deformable alignment~\cite{deformv2} in multi-scale feature space rather than pixel domain warping. \\
2) We propose multi-scale conditional B-frame encoding/ decoding for both offset and residual compression enabling recovery of both coarse and fine details at the decoder side utilizing a single compressive bottleneck. \\
3) A temporal latent extractor (TLE) is employed in addition to the hyper-prior to capture any relevant temporal information for entropy modeling. In addition, a parallel spatial-channel context model is also used to benefit from spatial and inter-channel correlations for bitrate reduction. \\
4) We employ content-adaptive video encoding by allowing online updating of the video encoder during inference. \\
5) We adopt the gain unit~\cite{Cui_2021_CVPR} to achieve the state-of-the-art multi-rate video compression results using a single model. 
\vspace{-4pt}

\begin{figure*}[ht]
\centering
	\includegraphics[width=0.95\textwidth]{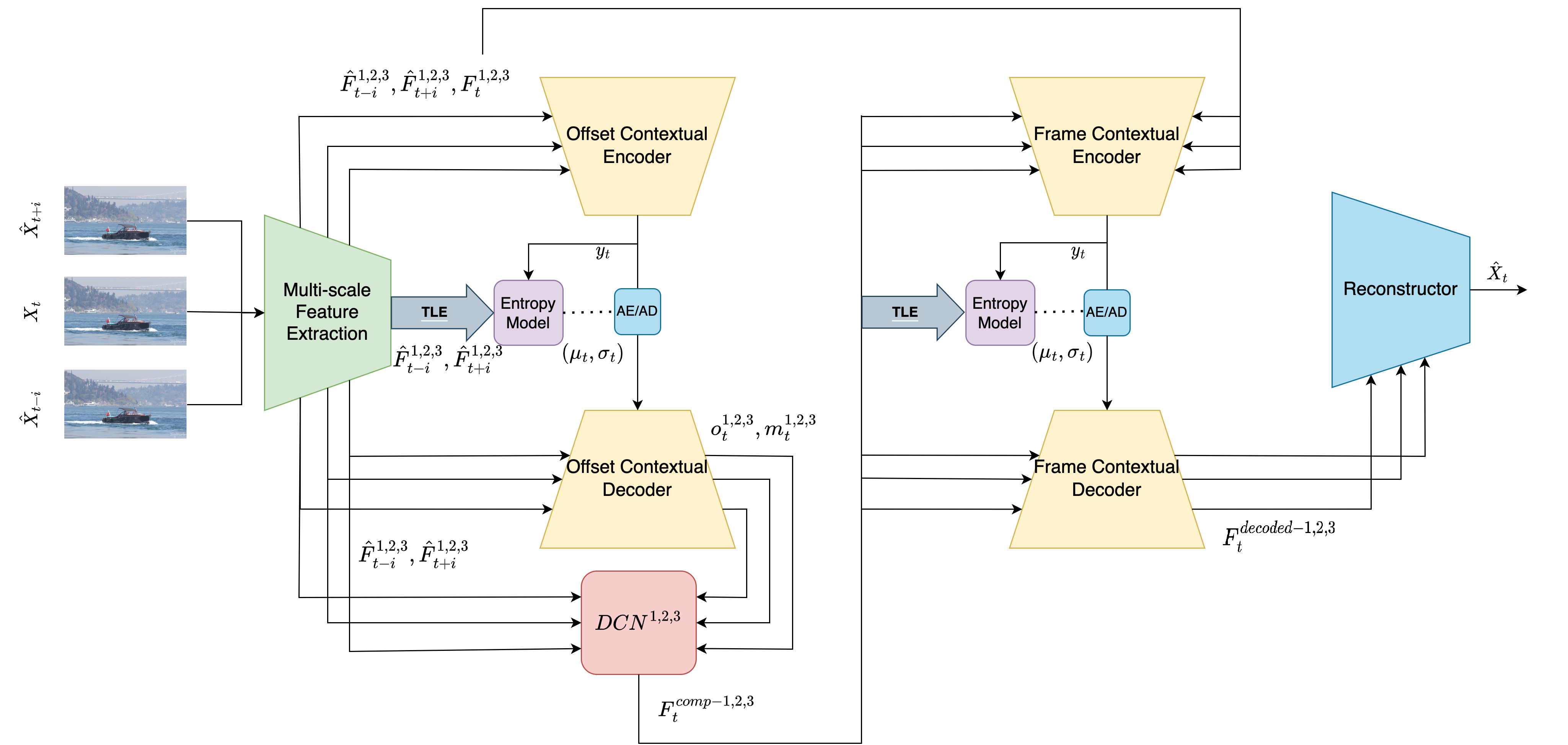} \vspace{-8pt} \\
\caption{Overview of the proposed bi-directional (B-frame) compression model. $t$ and $i$ denote the current frame  and past/future reference frame distance, respectively. In our design, we set the GoP size to 16, so $i$ can take a value from the list (1, 2, 4, 8).}
\label{fig:framework}
\end{figure*}

\section{METHOD}
\label{Method} \vspace{-4pt}
We extend deformable alignment and coding in the feature space for sequential video coding~\cite{dcvc,ZH_CVPR22} to multi-scale implicit motion compensation and conditional coding of B-frames. The overview of the proposed codec is shown in Figure~\ref{fig:framework}. \vspace{-18pt}
\subsection{Multi-scale Deformable Feature-Level Alignment}
\label{deform_align} \vspace{-4pt}
Given the current frame to be coded $x_{t}$ and past and future decoded reference frames $x_{t-i}$ and $x_{t+i}$, respectively, a shared feature extraction module computes multi-scale feature maps $\hat{F}^{1,2,3}_{t-i}$,
$\hat{F}^{1,2,3}_{t+i}$, $F^{1,2,3}_{t}$ for each frame separately. The feature extraction module, consisting of multiple strided convolutions and residual blocks, maps RGB frames with size $3 \times H \times W$ to 3 levels of feature tensors having sizes $32 \times H/2 \times W/2$, $64 \times H/4 \times W/4$ and $96 \times H/8 \times W/8$, respectively.

Next, a contextual encoder takes the feature tensors $\hat{F}^{1,2,3}_{t-i}$, $\hat{F}^{1,2,3}_{t+i}$, $F^{1,2,3}_{t}$ and maps them to latent variables~$y_{t}$. The contextual decoder predicts multi-scale offsets $o^{1,2,3}_{t}$ and modulation masks $m^{1,2,3}_{t}$ to be used in multi-scale deformable alignment by a deformable convolution net (DCN). 

Three different deformable convolution operations are applied at each level to get coarse to fine detail aligned features $F^{comp-1,2,3}_{t}$ at time $t$. In order to effectively use the complementary information, the past and future feature maps are concatenated as input to deformable convolutions as
\begin{equation}
F^{comp-1,2,3}_{t} = Dcn^{1,2,3}(c(\hat{F}^{1,2,3}_{t-i}, \hat{F}^{1,2,3}_{t+i}), o^{1,2,3}_{t}, m^{1,2,3}_{t})  \label{eq:deformconv} \vspace{-1pt}
\end{equation}
where $Dcn^{1,2,3}$ denotes deformable convolution layers and $o^{1,2,3}_{t}, m^{1,2,3}_{t}$ are the offsets and modulation masks, respectively. Note that the feature maps $\hat{F}^{1,2,3}_{t-i}$, $\hat{F}^{1,2,3}_{t+i}$ and aligned features $F^{comp-1,2,3}_{t}$ can be identically computed at both the~encoder and decoder (from the received bits) sides.
 \vspace{-2pt}
 
\subsection{Multi-scale Conditional Coding}
\label{cond_coding}
At the encoder, we encode and decode multi-scale feature map $F^{1,2,3}_{t}$ via conditional coding using $F^{comp-1,2,3}_{t}$ as condition using a single bottleneck as \vspace{-2pt}
 \begin{align}
\label{eq:encdec}
    & \hat{y}_{t} = round(Enc(F^{1,2,3}_{t},F^{comp-1,2,3}_{t})\\
    & F^{decoded-1,2,3}_{t} = Dec(\hat{y}_{t}, F^{comp-1,2,3}_{t})
\end{align}
 which was shown to be effective for sequential coding in~\cite{hybridspatio}.
 
Given the decoded feature maps $F^{decoded-1,2,3}_{t}$ at 3 different scales, we reconstruct the current decoded frame using a Unet architecture:
\begin{align}
\label{eq:reconst}
\hat{x}_{t} = Rec(F^{decoded-1}_{t}, F^{decoded-2}_{t}, F^{decoded-3}_{t})
\end{align}

\subsection{Spatio-Temporal Entropy Modeling}
\label{ent_model} \vspace{-4pt}
In order to model the entropy of latent space tensors $(y_{t})$ of both the offset and frame compression bottlenecks, hyper-priors of the respective latent spaces are first computed. Then a Temporal Latent Extractor module (TLE) maps the information available at the decoder side to a latent space tensor for better estimating a conditional entropy model. The TLE module has the same architecture as the Offset and Frame Contextual Encoders not using any information about the current frame to be compressed. Like~\cite{hybridspatio}, we fuse hyper-prior and TLE extracted latents via multiple convolution and residual blocks. We model each symbol $y^{i,j,k}_{t}$ by a Gaussian distribution where $(i, j, k)$ are the spatial and channel coordinates, respectively. It was shown that spatial and channel context models significantly improve the coding performance since a better entropy model can be estimated. We follow the work in~\cite{elic} and apply a parallel backward adaptive spatial and channel-wise context model for entropy modeling. 
\vspace{-4pt}

\subsection{Variable-Rate Compression}
\label{gain_unit} \vspace{-4pt}
For a practical video codec, flexible rate compression support is crucial. We employ the "gain" and "inverse gain" units explained in~\cite{Cui_2021_CVPR} to achieve variable rate coding. As the latent representations for the offset residuals and frame residuals both with $N=128$ channels, we scale the latent representations with two separate vectors of length $L=128$ in the channel dimension. This way, we effectively change the quantization bin sizes before rounding the latent representation to the nearest integer at inference time.

\subsection{Content-Adaptive Inference}
\label{content_adaptive}
\vspace{-6pt}
We propose a one-shot tuning of the learned model for adaptation to video type, resolution, and motion range. Traditional video codecs perform optimization customized to specific video content during encoding; thus, they can provide content adaptation for increased coding efficiency. Following~\cite{contentadap}, we employ an online encoder update strategy during inference to better handle different types of video content. 
\vspace{-6pt}

\begin{figure}[b!]
\centering
	\includegraphics[scale=0.33]{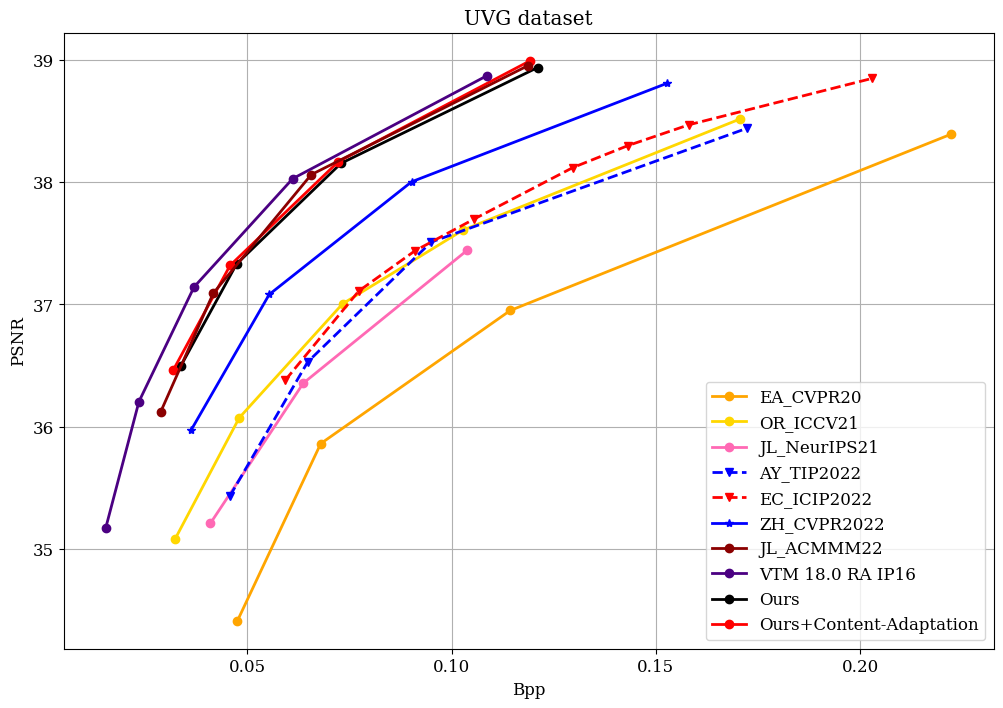} \vspace{-6pt} \\
\caption{Comparison of R-D performance of the proposed B-frame compression model with prior works in terms of PSNR.}
\label{fig:psnr_curve} \vspace{-14pt}
\end{figure}

\section{Experimental Results}
\label{eval}
\vspace{-5pt}
\subsection{Datasets, Training and Implementation Details} \vspace{-4pt}
We used Vimeo90k septuplet dataset~\cite{vimeo} for end-to-end training of the model. For testing, we evaluate the performance of our method on widely used 1080p UVG dataset~\cite{uvg}.

For intra frame compression, we train our own image compression model for 4 different compression levels following\cite{elic}. We then train our B-frame compression model for 4~quality levels using Adam optimizer~\cite{adam}. We pair intra and B-frame quality levels for rate adaptation. 2M optimization steps are performed to minimize the rate-distortion loss. For the first 750K iterations, we randomly select 3 consecutive frames and compress the middle frame given future and past reference frames. After 750K steps of optimization, 5 consecutive frames are randomly selected and loss is propagated for 2 hierarchy levels of compression to better handle error propagated through coding order. During training, the batch size is set to 4. Random cropping of $256 \times 256$ pixels and random temporal flipping are applied for data augmentation. Learning rate of the optimizer is set to 1e-4 at the beginning and halved if validation loss stays stable for 100K iterations. We applied gradient norm clipping for stable training. 
\vspace{-4pt}

\subsection{Encoder Fine-tuning Strategy for Content Adaption}
\vspace{-4pt}
We freeze all components of the proposed model except for the offset and frame compression encoders and hyper-encoders. We optimize encoder parts for each frame to be coded using the Adam optimizer. We set the learning rate to 1e-5 and perform optimization for 100 steps at maximum. If~the rate-distortion loss of a frame does not decrease for 10~steps, we apply early-stopping. 
\vspace{-4pt}

\subsection{Comparison with the State-of-the-Art}\vspace{-4pt}
Rate-distortion performances of our method with content-adaptive inference and previous state-of-the-art learned codecs as well as the VTM-18.0 codec are shown on Figure\ref{fig:psnr_curve}. The~quantitative comparisons are presented in terms of peak-signal-to-noise-ratio (PSNR)~\cite{psnr_comp} and bits-per-pixel (bpp) averaged across all frames from all sequences. It can be clearly seen that our method indubitably outperforms other learned frameworks except we achieve comparable results with the latest state-of-the-art codec~\cite{hybridspatio}. 
We also checked our performance on MCL-JCV dataset~\cite{mcljcv} and confirmed that our model generalizes well on a diverse set of videos. RD points and curves are available on the project web page. 

The average BD-BR improvements vs. the VTM-18.0 encoder over the 7 UVG sequences are presented on the right hand side of Figure\ref{fig:rate_savings}. On average, our model without content-adaptation is on par with the~\cite{hybridspatio} and some improvement can be observed when we perform content-adaptive inference. Figure\ref{fig:rate_savings} also shows results for individual sequences \textit{Beauty}, \textit{Honeybee}, and \textit{YatchRide}.
On the \textit{Honeybee} sequence, which contains occluded regions, our performance significantly exceeds that of the state of the art learned sequential coding model~\cite{hybridspatio} underlining the power of bi-directional compression. 
Indeed, for all the sequences except \textit{Jockey} and \textit{Ready}, which contain fast and complex motions, our method achieves superior results compared to~\cite{hybridspatio} and our results are better or on par with the VTM-18.0. 
\vspace{-4pt}

\subsection{Ablation: Content Adaptation vs. No adaptation}
\begin{figure}[t]
\centering
	\includegraphics[scale=0.20]{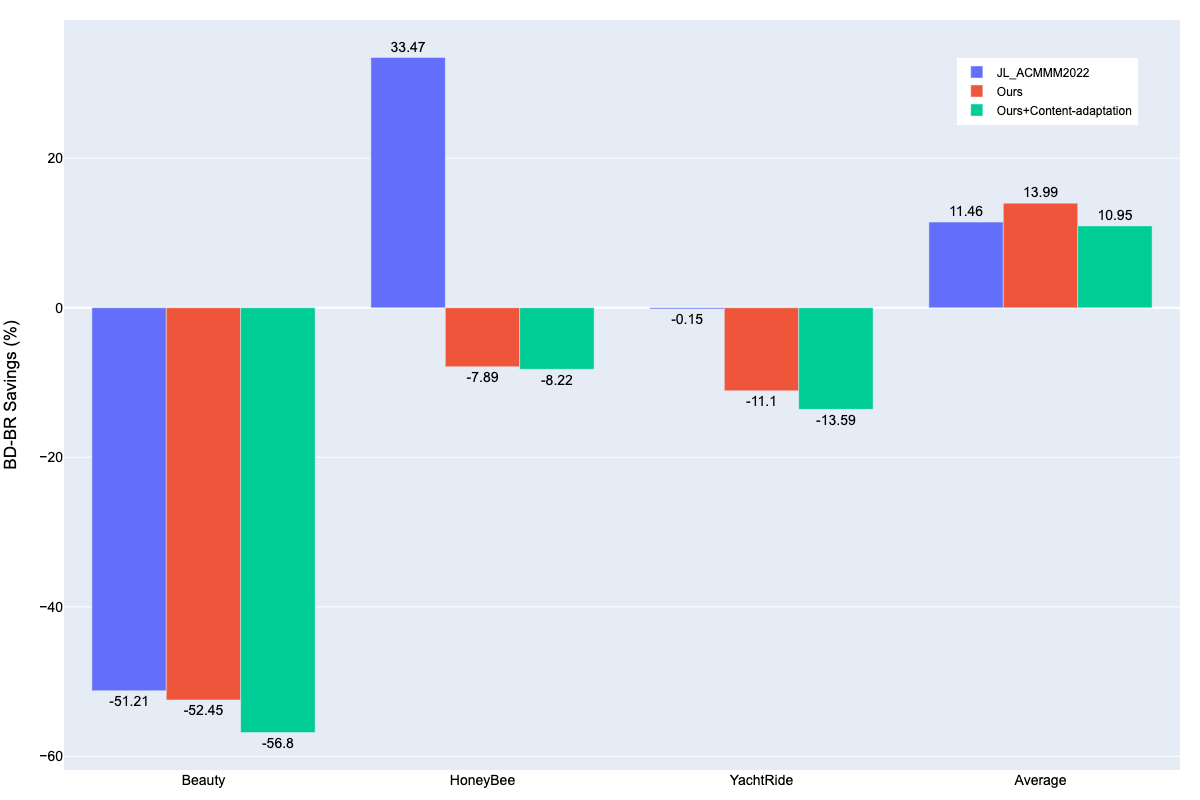} \vspace{-6pt} \\
\caption{Average percent BD-BR improvements (RGB bpp)
for the proposed model and JL\_ACMMM22~\cite{hybridspatio} vs. the anchor VTM 18.0 encoder (RA IP16) on the UVG sequences.}
\label{fig:rate_savings}
\end{figure}

\vspace{-4pt}
In Figure\ref{fig:rate_savings}, we present both content-adaptive and non-adaptive inference results. Our results reveal that content-adaptation provides better performance. However, the benefits of content adaptation may vary depending on the complexity of the motion in the video. In our experiments, we observed that content adaptation was particularly effective for sequences with complex motion patterns that were not well-represented in our training data. For such cases, our method was able to achieve significant gains in BD-BR, indicating that content adaptation can help bridge the gap between training and test data distributions. On the other hand, for sequences that were already well-represented in our training data and had simple motion patterns, the gains from content adaptation were smaller, but still noticeable. This suggests that content adaptation can be a useful tool for improving video compression in a wide range of scenarios, particularly for challenging sequences that may not be adequately captured by existing compression techniques.

\section{Conclusion}
\label{conc}
\vspace{-3pt}
We propose an end-to-end optimized multi-scale deformable alignment and conditional coding framework that is trained once to operate at multiple bitrates. We also applied online encoder updating strategy for content-adaptive optimization during inference. The resulting model yields superior R-D performance in terms of PSNR vs. RGB bits-per-pixel. 

Our model provides improvements by employing deformable alignment and conditional coding at multiple resolutions while supporting variable rate coding via a single trained model. We also investigated online encoder optimization strategy and found that it brings noticeable performance improvements especially for complex sequences. 

We plan to further investigate better modeling of complex motions as future work.


\clearpage
\bibliography{references}

\begin{thebibliography}{10}
\providecommand{\url}[1]{#1}
\csname url@samestyle\endcsname
\providecommand{\newblock}{\relax}
\providecommand{\bibinfo}[2]{#2}
\providecommand{\BIBentrySTDinterwordspacing}{\spaceskip=0pt\relax}
\providecommand{\BIBentryALTinterwordstretchfactor}{4}
\providecommand{\BIBentryALTinterwordspacing}{\spaceskip=\fontdimen2\font plus
\BIBentryALTinterwordstretchfactor\fontdimen3\font minus
  \fontdimen4\font\relax}
\providecommand{\BIBforeignlanguage}[2]{{%
\expandafter\ifx\csname l@#1\endcsname\relax
\typeout{** WARNING: IEEEtran.bst: No hyphenation pattern has been}%
\typeout{** loaded for the language `#1'. Using the pattern for}%
\typeout{** the default language instead.}%
\else
\language=\csname l@#1\endcsname
\fi
#2}}
\providecommand{\BIBdecl}{\relax}
\BIBdecl

\bibitem{h265}
G.~J. {Sullivan}, J.~{Ohm}, W.~{Han}, and T.~{Wiegand}, ``Overview of the high
  efficiency video coding (hevc) standard,'' \emph{IEEE Trans. on Circuits and
  Systems for Video Tech.}, vol.~22, no.~12, pp. 1649--1668, Dec 2012.

\bibitem{vvc}
B.~Bross, Y.-K. Wang, Y.~Ye, S.~Liu, J.~Chen, G.~J. Sullivan, and J.-R. Ohm,
  ``Overview of the versatile video coding (vvc) standard and its
  applications,'' \emph{IEEE Transactions on Circuits and Systems for Video
  Technology}, vol.~31, no.~10, pp. 3736--3764, 2021.

\bibitem{vtm}
B.~Bross, J.~Chen, S.~Liu, and Y.-K. Wang, ``Versatile video coding (draft
  10),'' \emph{Joint Video Experts Team (JVET) of ITU-T SG16 WP3 and ISO/IEC
  JTC 1/SC29, Output Document JVET-S2001}, 2020.

\bibitem{agustsson_scale}
E.~Agustsson, D.~Minnen, N.~Johnston, J.~Ballé, S.~J. Hwang, and G.~Toderici,
  ``Scale-space flow for end-to-end optimized video compression,'' in
  \emph{2020 IEEE/CVF Conference on Computer Vision and Pattern Recognition
  (CVPR)}, 2020, pp. 8500--8509.

\bibitem{elfvc}
O.~Rippel, A.~G. Anderson, K.~Tatwawadi, S.~Nair, C.~Lytle, and L.~Bourdev,
  ``Elf-vc: Efficient learned flexible-rate video coding,'' \emph{2021 IEEE/CVF
  International Conference on Computer Vision (ICCV)}, 2021.

\bibitem{ladune2021conditional}
T.~Ladune, P.~Philippe, W.~Hamidouche, L.~Zhang, and O.~D{\'e}forges,
  ``Conditional coding for flexible learned video compression,'' in
  \emph{Neural Compression: From Information Theory to Applications -- ICLR
  Workshop}, 2021.

\bibitem{dcvc}
J.~Li, B.~Li, and Y.~Lu, ``Deep contextual video compression,'' \emph{Advances
  in Neural Information Processing Systems}, vol.~34, 2021.

\bibitem{ZH_CVPR22}
Z.~Hu, G.~Lu, J.~Guo, S.~Liu, W.~Jiang, and D.~Xu, ``Coarse-to-fine deep video
  coding with hyperprior-guided mode prediction,'' in \emph{2022 IEEE/CVF
  Conference on Computer Vision and Pattern Recognition (CVPR)}, 2022, pp.
  5911--5920.

\bibitem{hlvc}
R.~Yang, F.~Mentzer, L.~Van~Gool, and R.~Timofte, ``Learning for video
  compression with hierarchical quality and recurrent enhancement,'' in
  \emph{IEEE/CVF Conf. on Computer Vision and Patt. Recog. (CVPR)}, 2020.

\bibitem{lhbdc}
M.~A. Yılmaz and A.~M. Tekalp, ``End-to-end rate-distortion optimized learned
  hierarchical bi-directional video compression,'' \emph{IEEE Transactions on
  Image Processing}, vol.~31, pp. 974--983, 2022.

\bibitem{flexrate}
E.~Çetin, M.~A. Yılmaz, and A.~M. Tekalp, ``Flexible-rate learned
  hierarchical bi-directional video compression with motion refinement and
  frame-level bit allocation,'' in \emph{2022 IEEE International Conference on
  Image Processing (ICIP)}, 2022, pp. 1206--1210.

\bibitem{Cui_2021_CVPR}
Z.~Cui, J.~Wang, S.~Gao, T.~Guo, Y.~Feng, and B.~Bai, ``Asymmetric gained deep
  image compression with continuous rate adaptation,'' in \emph{IEEE/CVF Conf.
  on Computer Vision and Pattern Recognition (CVPR)}, June 2021, pp.
  10\,532--10\,541.

\bibitem{deformv2}
X.~Zhu, H.~Hu, S.~Lin, and J.~Dai, ``Deformable convnets v2: More deformable,
  better results,'' \emph{2019 IEEE/CVF Conference on Computer Vision and
  Pattern Recognition (CVPR)}, pp. 9300--9308, 2018.

\bibitem{hybridspatio}
J.~Li, B.~Li, and Y.~Lu, ``Hybrid spatial-temporal entropy modelling for neural
  video compression,'' in \emph{Proceedings of the 30th ACM International
  Conference on Multimedia}, 2022.

\bibitem{elic}
D.~He, Z.~Yang, W.~Peng, R.~Ma, H.~Qin, and Y.~Wang, ``Elic: Efficient learned
  image compression with unevenly grouped space-channel contextual adaptive
  coding,'' in \emph{IEEE/CVF Conference on Computer Vision and Pattern
  Recognition (CVPR)}, 2022, pp. 5718--5727.

\bibitem{contentadap}
G.~Lu, C.~Cai, X.~Zhang, L.~Chen, W.~Ouyang, D.~Xu, and Z.~Gao, ``Content
  adaptive and error propagation aware deep video compression,'' in
  \emph{Computer Vision -- ECCV 2020}, 2020, pp. 456--472.

\bibitem{vimeo}
T.~Xue, B.~Chen, J.~Wu, D.~Wei, and W.~T. Freeman, ``Video enhancement with
  task-oriented flow,'' \emph{International Journal of Computer Vision (IJCV)},
  vol. 127, no.~8, pp. 1106--1125, 2019.

\bibitem{uvg}
A.~Mercat, M.~Viitanen, and J.~Vanne, ``Uvg dataset: 50/120fps 4k sequences for
  video codec analysis and development,'' in \emph{ACM Multimedia Systems
  Conference}, ser. MMSys '20, 2020, p. 297–302.

\bibitem{adam}
D.~P. Kingma and J.~Ba, ``Adam: A method for stochastic optimization,'' in
  \emph{Int. Conf. Learning Representation (ICLR)}, 2015.

\bibitem{psnr_comp}
O.~Keleş, M.~A. Yilmaz, A.~M. Tekalp, C.~Korkmaz, and Z.~Doğan, ``On the
  computation of psnr for a set of images or video,'' \emph{2021 Picture Coding
  Symposium (PCS)}, pp. 1--5, 2021.

\bibitem{mcljcv}
H.~Wang, W.~Gan, S.~Hu, J.~Y. Lin, L.~Jin, L.~Song, P.~Wang, I.~Katsavounidis,
  A.~Aaron, and C.-C.~J. Kuo, ``Mcl-jcv: A jnd-based h.264/avc video quality
  assessment dataset,'' in \emph{2016 IEEE International Conference on Image
  Processing (ICIP)}, 2016, pp. 1509--1513.

\end{thebibliography}
\bibliographystyle{IEEEtran}
\end{document}